\documentclass[12pt, titlepage]{article}
\usepackage{amsmath,amssymb}
\begin{document}

{\center
%\begin{titlepage}   
%\title
{   \Large \bf{Osmotic Properties of Charged Cylinders:
Critical Evaluation of Counterion  Condensation Theory} } 
\\[5mm]
{\large Per Lyngs Hansen,
Rudi Podgornik $\S$, V. Adrian Parsegian  \\[3mm] 
{Laboratory of Physical and Structural Biology} \\
{National Institute of Child Health and Human Development} \\
{National Institutes of Health, Bethesda, MD 20892-5626} \\
{$\S$ Department of Physics} \\
{Faculty of Mathematics and Physics} \\
{University of Ljubljana, SI-1000 Ljubljana, Slovenia} \\[5mm]
}
 } 

\date{\today}
%\maketitle
%\begin{abstract}
{\small
\noindent

\vskip 2 cm 
The osmotic coefficient of B-DNA in water may, in moderately dilute
solutions, deviate as much as 100 \% from predictions based on a
simple 'counterion condensation' theory.  We determine the results for
osmotic properties via a cell model description of the ionic
atmosphere near a cylindrical polyelectrolyte.  The cell model
predictions for the osmotic properties disagree with predictions based
on simple condensation theory, but are in surprisingly good harmony
with experimental findings.  We argue that the neglect of
finite-radius effects makes simple condensation theory inapplicable at
all but impractically low polyelectrolyte concentrations and, unable to
reproduce osmotic properties of polyelectrolytes such as DNA. }

\eject

%On the other hand, in the line-charge 
%limit simple condensation theory provides a perfectly valid linearized 
%description of polyelectrolyte properties.

%In the analysis of the 
%differences between the cell model approach and simple condensation theory 
%we highlight that simple condensation theory may be thought of as a limiting case of the cell
%model description when the radius of the cyldinder is sent to zero, 
%but is otherwise in disagreement with the cell model analysis. 
%The inabilty
%of simple condensation theory to account for osmotic properties of DNA 
%seems therefore to be  due to its neglect of finite-radius effects.

%                       *********************
%                           {Introduction}

\vspace{1cm}
\noindent
Ionic screening of charge interactions remains one of the most
vigorously discussed properties of
%The question of local aspects of screening remains one of the   
%important topics in  the discussion of  properties of
polyelectrolyte solutions \cite{Joanny,Frank-Kamenetskii}.  Recently
measured osmotic coefficients of B-DNA in dilute aqueous solutions
\cite{Livolant}, show factor-of-two deviations from predictions based
on the influential counterion condensation theory of Oosawa and
Manning \cite{Oosawa,Manning69, Manning98}.  We shall argue that an
earlier theory pioneered by Lifson and Katchalsky
\cite{Lifson,LeBret}, based on the cell model formulation of the full
nonlinear Poisson-Boltzmann equation , provides a more successful
starting point than simple condensation theory for examining not only
the osmotic properties of cylindrical polyelectrolytes in low salt
conditions but also the local organization of the ionic atmosphere
around them.  By explicitly including the finite size of the cylinder,
the cell model description fundamentally disagrees with the line
charge picture of the simplest condensation theory.  It is well known
\cite{Oosawa, Manning69} that in the absence of added salt, the simple
condensation theory is a special limiting case of the cell model only
when the cylinder radius is sent to zero or the cell radius is sent to
infinity.  We find here that neglect of finite radius is the main
reason for the failure to describe DNA osmotic properties.

%yet (obviously) also shows that in the appropriate geometric limit 
%simple condensation theory will be a useful and quite correct effective 
%linearized description of polyelectrolyte solution properties.

%ADRIAN'S stuff
%This is not a general criticism of ion condensation theory and its many
%reported successes. It is to point out that in 2 mM salt solutions the
%osmotic pressure of charged hollow rods calculated by the PB equation as
%well as the directly measured the pressure are both about twice that
%predicted by condensation theory.  It is also to point out that the osmotic
%coefficient in experiment, and to a greater extent in PB computation,
%varies slightly with polyelectrolyte concentration while ion condensation
%theory, by its very construction, predicts a constant osmotic coefficient. 

%Finite radius seems to be the culprit. As is well known, when a/R à 0, the
%P-B solution goes uniformly to the ion-condensation line-charge limit. Our
%calculations suggest that 2 mM solutions even for a/R ~.1 are still much
%too large to allow us to assume behavior at the line charge limit. 

%The observed P-B predicted osmotic pressures are always much greater that
%the osmotic pressure of a 2 mM salt solution. Therefore use of the
%salt-free Lifson, Katchalsky formula is appropriate. 

%RUDI'S stuff
The osmotic pressure of a polyelectrolyte (e.g., DNA) in an aqueous
solution is obtained from the equation of state that recognizes the
degrees of freedom of both the counterion and the macroion.  The
Oosawa-Manning limit \cite{Oosawa, Manning69} which decouples the
counterion atmosphere from the density of the macroion can thus only
make sense at effectively infinite dilution.  Osmotic pressure
experiments however are performed at finite macroion concentrations. 
In these experiments \cite{Livolant} the osmotic coefficient of DNA in
2 and 10 mM salt solutions is seen to be twice that predicted by the
counterion condensation theory, while the Lifson - Katchalsky cell
model accurately predicts its magnitude.  The experimental osmotic
coefficient varies weakly with the macroion concentration in the
general direction predicted by the cell model.  

%The most drastic prediction is
%that the osmotic coefficient is about twice the value predicted by
%Oosawa-Manning theory.  For this particular experiment, the Oosawa-Manning
%is thus seen to be too stringent and does not fully account for the
%macroion density dependence.

%                         *********************
%                                {Model}
          
\vspace{1cm}
\noindent
%In order to establish these conclusions let us recall that in the absence
%of salt, 
The cell model considered here involves a rigid (hollow or solid),
charged cylindrical polymer of radius $a$, coaxially enclosed in a
cylindrical (Wigner-Seitz-like) cell of radius $R_{0}$, corresponding
to the total system volume per polymer length, see Fig.~1.  The cell acts as a
neutralization volume for the counterions; consequently the electric
field vanishes at the cell wall.  Counterions organize within the cell
according to the nonlinear Poisson-Boltzmann equation for the double
layer electric potential, $u$.  In units of $k_{B}T/e$, in the absence
of added salt,
\begin{eqnarray}
\frac{1}{r}\frac{d}{dr}(r\frac{d u_{*}(r)}{dr}) = -\frac{{\kappa_{*}}^{2}}{2} 
e^{{-u_{*}(r)}}~. \label{PB}
\end{eqnarray}
Here $\kappa^{-1}_{*}$ denotes a formal 'screening length' to be
determined later by the average density of counterions.  The subscript
$*$ stands for the electrostatic potential inside and outside of the
cylinder The potentials in turn determine the charge densities,
\begin{eqnarray}
n_{*}(r) = n_{*,0} e^{-u_{*}(r)}~. \label{Boltzmann}
\end{eqnarray}
Here  $n_{*,0}=\kappa^{2}_{*}/(8\pi l_{Bj})$, and $l_{Bj}=e^{2}/(\epsilon k_{B}T)$ 
is the Bjerrum length. 

Because of the major and minor grooves in DNA \cite{Saenger}, it is
desireable to treat the cylinder at least as partly hollow and hence to
give solvent and ions access to the space within the grooves.  We
shall compute here, for simplicity, the results for cylinders that are
either solid or completely hollow to make the maximum possible
'non-specific' accumulation of countercharge near the cylinder. 
Counterion accumulation is determined by the solutions to the PB
equation for which the density variation across the boundary at $r=a$
is continuous.
%In practice it is convenient to also impose that the potential $u$
%varies continuously across the boundary, $r=a$. This may be achieved 
%by imposing that 
$u_{o} (r\geq a)$  and $u_{i} (r\leq a)$ are both expressed with respect to
a zero at the cell wall ($r=R_{0}$) and $\kappa^{2}_{i}=\kappa^{2}_{o} 
=8\pi l_{Bj}n(R_{0})$, where $n(R_{0})$ is the density of counterions 
at the cell wall. 

%The relevant solutions to Eq.~(\ref{PB}) are, in fact, well-known:  
For $a< r < R_{0}$ solution of the Poisson - Boltzmann equation yields
\cite{Lifson}
\begin{eqnarray}
u_{o}(r) = 
\ln\left(~\frac{{(\kappa r)}^{2}}{2z}\cos^{2}(2\ln(\frac{r}{R_{m}}))~\right)~,   
\end{eqnarray}
 \cite{Lifson,LeBret} and for $r < a$ one finds  \cite{Parsegian}
\begin{eqnarray} 
u_{i}(r) = u_{0}+ 2 \ln\left(1 + c {r}^{2}\right)~.
\end{eqnarray} 

%The choice $u(R_{0})=0$ and boundary conditions yield the equations 
The integration constants $z, R_{m},\kappa$, $c$, and $u_{0}$ are
obtained from boundary conditions.  In addition to the requirement of
a continuous variation of the potential at $r=a$, these conditions
include $du_{i}(r)/dr|_{0} = 0$, $du_{o}/dr|_{R_{0}}=0$, and
$(du_{o}(r)/dr - du_{i}(r)/dr)|_{a}=2Q/a$.  For DNA the dimensionless
linear charge density $Q=l_{Bj}/l_{P04}\simeq 4.353$ is determined by
the charge separation $l_{PO4}\simeq 1.7 \AA$ and the Bjerrum length
$l_{Bj}\simeq 7.14 \AA$ (as in water at room temperature).  In practice
one solves the problem of fixing the parameters by an iteration 
that starts with a trial partitioning of counterions inside and
outside the cylinder that is subsequently refined until the
boundary conditions are exactly satisfied.

%one picks trial values of $Q_{i}$ and $Q_{o}$,
%that in their turn determine the number of counterions inside and outside
%the cylinder.  This fixes unique values of ($c$, $u_{0}$), and ($z$,
%$R_{m}$, $\kappa$), respectively.  The physically acceptable solution,
%which may be obtained by iteration, is the one for which $Q_{i}+Q_{o}=Q$

%                         ************************ 
%                                 {Results}

\vspace{1cm} \noindent The osmotic properties of charged cylinders are
encoded in the osmotic pressure $\pi_{osm}=k_{B}T n(R_{0})$ on the
cell wall, which, of course, also codifies the equation of state for
the macroion at the density set by $R_{0}$.  (Because the electric
field vanishes at the cell wall, the only contribution to the Maxwell
stress tensor comes from the osmotic pressure.)  The osmotic
coefficient $\phi$ is defined as the ratio of the actual osmotic
pressure to the osmotic pressure of a hypothetical gas of uniformly
distributed counterions or, equivalently, the number density of ions
$n(R_{0})$ divided by the total density $n_{PO4} = 1/(l_{PO4}\pi
R_{0}^{2})$: .

Fig.~1.  shows the results of a simple numerical calculation of the osmotic
coefficient of solutions of rigid hollow cylinders as a function of the
molar concentration, when the radius of the cylinder $a=10\AA$, the
dimensionless (nominal) line charge density $Q=4.353$, and the molar weight
$M$ is chosen as for DNA.

{\it (i)} The calculated osmotic coefficients vary slightly with
concentration.  The measured osmotic coefficients
\cite{Livolant} vary even more slowly.  However, simple condensation theory
predicts no variation at all.  

{\it (ii)} In the region of significant experimental interest, i.e.,for
concentrations of DNA phosphates from $0.1$ to $0.5$ M, the calculated osmotic
coefficients are close to $0.3$ in reasonably close agreement with the
experimental value $0.24$ \cite{Livolant}.  Simple condensation
theory predicts $\phi=1/(2Q)\simeq 0.11$.  

{\it (iii)}As expected in the limit $a/R_{0}\longrightarrow 0$, our
formulated osmotic coefficients converge to the line - charge result
$\phi(a/R_{0}) \longrightarrow 1/(2Q)$; it is the significant
deviations that occur away from this infinite dilution limit that
concern us.

Specifically, consider the asymptotic expression for the osmotic coefficient in the
limit where $a/R_{0}$ is small (the line-charge limit, or the 
infinite-dilution limit): The outer solution with
\begin{eqnarray}
Q_{o}= 1- z\tan\left(z\ \ln(\frac{a}{R_{m}}) \right) \\ \label{BC1}
0= 1- z\tan\left(z\ \ln(\frac{R_{0}}{R_{m}}) \right) \\ 
\kappa^{2}R_{0}^{2}=4(1+z^{2})~.
\end{eqnarray}
give us $z$, as follows:
\begin{eqnarray}
\ln(\frac{a}{R_{0}}) = \frac{\arctan\left(\frac{1-Q_{o}}{z}\right) 
                        - \arctan\left(\frac{1}{z}\right)}{z}~. 
\end{eqnarray}
When $a/R_{0}$ is small, $z 
\longrightarrow 0$, the $\arctan(\cdot)$'s may then be replaced by, respectively,
$-\pi/2$ and $\pi/2$ so that z has a weak logarithmic 
dependence on $a/R_{0}$:
\begin{eqnarray}
 z\simeq \frac{\pi}{ \ln(\frac{R_{0}}{a}) }~. \label{BC2} 
\end{eqnarray}
The corresponding asymptotic expression for the osmotic coefficient is 
\begin{eqnarray}
\phi =\frac{n(R_{0})}{n_{PO4}}=\frac{\kappa^{2}}{8\pi 
l_{Bj}}\pi R_{0}^{2}l_{PO4}  = 
\frac{1}{2Q_{o}}(1+z^{2}) G(a/R_{0}) ~,
\end{eqnarray}
where $G(a/R_{0})$ is a geometric factor, which for hollow cylinders
assumes the value 1. Therefore, for small values of $a/R_{0}$,
Eq. (\ref{BC2}) gives 
\begin{eqnarray}
\phi \simeq 
\frac{1}{2Q_{o}}\left(1+\frac{\pi^{2}}{\ln^{2}(\frac{R_{0}}{a})} \right) ~.
\end{eqnarray}
In other words, the osmotic coefficient develops a weak logarithmic
concentration dependence.  (If the cylinders were not completely
hollow an extra concentration dependence would appear at high
concentrations.  In this case the geometric factor $G(a/R_{0})$ would
deviate from unity.)  Any concentration dependence of the osmotic
coefficient is incompatible with simple condensation theory.  Only as
$a/R_{0}\longrightarrow 0$ does the osmotic coefficient approach a
concentration-independent limit $1/(2Q_{o})$. 
%$a/R_{0}\longrightarrow 0$.
%where $z\longrightarrow 0$ and $Q_{o}\longrightarrow Q$. 

\vspace{1cm}
\noindent 
%The comparison of cell model calculations and experimental data
%suggest that the osmotic properties are dominated by the
%electrosatic component. In order to learn more about this topic we 
%have computed the cell model predictions for the osmotic pressure
%and compared the results with available data.

Fig.~2 shows the osmotic pressure ${\pi}_{osm}$ versus the molar
concentration $c_{DNA}$ of DNA or equivalently versus the radius of
the cell $R_0$.  Experimental data are available at low, $0.1-0.5$ M
\cite{Livolant} and high concentratios, $1-2$ M \cite{Helmut}.  In
both ranges we find that calculated osmotic pressures are remarkably
close to capturing the magnitude and variation of the measured
pressures.  In the low-concentration range and 2 mM salt the
calculated pressures are slightly too large which might in part
reflect a weak contribution from finite salt concentration that can be
dealt with on the basis of a simple Donnan equilibrium picture
\cite{Oosawa,Salt,Livolant}.  This approach however fails completely
at higher salts, e.g. 10 mM, where the osmotic pressure is
substabtially lower than the values calulated from the Donnan
equilibrium.  Though one could make the calculated values of osmotic
pressure in the intermediate regime of DNA densities to be even closer
to the data, by choosing a somewhat smaller value for the DNA radius,
it turns out that in this case one would loose the relatively good
agreement in the regime of very large DNA concentrations.  There
appears to be no simple way of adjusting the DNA parameters to get a
good quantitative fit of the calculated osmotic pressure with
experiments.  As pointed out many times before \cite{hydration} it
appears yet again that non-electrostatic interactions at very large
DNA concentrations make a significant contribution to the overall
osmotic pressure in the system.

In the high-concentration range, the calculated variation of the
pressure with concentration is clearly slower than what one observes
for the experimental data.  The (small) difference may be of
non-electrostatic origin \cite{hydration} or reflect
charge-discreteness effects discussed in Ref.  \cite{Sergey}: salt
effects are likely to be unimportant in this concentration range.  The
renormalization effects due to chain conformational fluctuations,
discussed in Ref.\cite{Helmut}, lead to predictions of decreasing
rates of change of the pressure with concentration.  Recent work on
stretching of DNA at various ionic conditions \cite{Bustamante} also
suggests that there might be an additional strong coupling between DNA
elasticity and electrostatics.  Though the details of this coupling
are only beginning to be elucidated \cite{Podgornik} it is concievable
that local deformations of DNA would change the countercharge
distributions at low salt conditions and thus effect also the osmotic
coefficient.

%Finally, it is quite possible that the single-molecule
%intrinsic stretchability which is considerable rather than negligable at
%low salt \cite{stretch} will be of relevance for estimate of the osmotic
%pressure of DNA solutions.
 
%                             ********************** 
%                                  (Discussion)

\vspace{1cm}
\noindent 
%Does condensation theory work ?
To what extent do the above observations elucidate 'counterion
condensation, 'simple' \cite{Oosawa,Manning69} or 'extended'
\cite{LeBret,Belloni,Holm} ?

%This might turn out to depend on what one means by condensation.
%We observe that it seems possible to introduce at least
%two concepts of counterion condensation, we might call them
%'simple' \cite{Oosawa,Manning69} or 'extended' \cite{LeBret,Belloni, 
%Holm} ?
% We shall argue here that the 'simple' condensation concept 
%is not coreect in the sense that it is not complete, whereas the 'extended' concept, 
%to the extent that  it describes condensation at all, may be argued to be 
%essentially correct. 

\vspace{1cm}
\noindent
% SImple condensation theory
In the 'simple condensation' picture \cite{Oosawa, Manning69} highly charged
and rigid cylindrical macromolecules are portrayed as line charges 
with explicit neglect of finite macromolecular radius, the atmosphere of
counterions is divided into a condensed and osmotically
inactive fraction which redefines the line charge density, and an
'unbound', osmotically active fraction.  This 'two-phase coexistence'
is established whenever the line-charge density (or, equivalently, the
electric field at the surface of the macromolecule) becomes large enough,
quantitatively whenever $Q=l_{Bj}/l_{PO4}> 1$,
%($l_{Bj}\simeq 7 \AA$ in water at room 
%temperature), and $b$ is the separation along the contour of the polymer
%between two elementary charges ($b\equiv l_{PO4} \simeq 1.7 \AA$ in DNA).
For $Q>1$, condensation will bring down the effective (dimensionless)
line-charge density from $Q$ to $1$; a fraction $f_{c}=(Q-1)/Q=1-1/Q$ of
the counterions condense.  The remaining fraction $f=1/Q$ of the charge
remains unbound.  If the unbound fraction is modeled as a polarized
Debye-H\'uckel gas of counterions, this gas contributes to the osmotic
coefficient $\phi$, as $\phi_{M}=\pi/(k_{B}T n_{Q})= f/2 =1/(2Q)$.
%in a way that only depends on the linear charge
%density, but not on the concentration of polymers. 
For DNA   $\phi_{M} \simeq 0.11$ while a significant fraction
$f \simeq 0.75$ of the countercharge is 'bound'. 

%This is the 'simple' picture of Oosawa and Manning. Various
%experimental works \cite{} all of which are conducted near the
%line-charge limit lends support to this picture. Ironically, perhaps, the cell 
%model also supports this picture in an important way. In fact, 
%following 
In the line-charge limit \cite{Oosawa, Manning69} there is no
disagreement at all between cell model predictions and predictions
based on 'simple' condensation theory.  In the cell model
\cite{LeBret} an effective 'condensation range' defines a Manning
radius $R_{M}$ such that the fraction of the countercharge contained
within a shell $a<r<R_{M}$ is $f=1-1/Q_{o}$, precisely the fraction of
charges that are predicted to condense according to the Oosawa -
Manning model.  Using Eqs.~(\ref{BC1})-(\ref{BC2}), in the small
$a/R_{0}$ limit, ${ R_{M}} \simeq { R_{0}} e^{-\frac{\pi}{2z}} \simeq
\sqrt{R_{0}a}$.  That is in this limit the Manning radius disappears;
the effects of the fraction $f=1-1/Q_{o}\longrightarrow 1-1/Q$ of
ionic atmosphere can be absorbed in a redefinition of the line charge,
as in 'simple' condensation theory.  However when $a > 0$ the
Manning radius ${ R_{M}}$ is finite and even diverges as one
approaches the infinite-dilution limit $R_{0}\longrightarrow \infty$.

Is it possible to reformulate or 'extend' condensation theory to
account for finite macromolecular radius while retaining the essential
idea of two-phase coexistence between free and bound fractions?  If
not, the organization of the counterion atmosphere around cylindrical
polyelectrolytes and the electrostatic contribution to solution
properties of polyelectrolytes, must be understood on the basis of a
Poisson-Boltzmann or a more advanced double layer description.  
To the extent that experiments like \cite{Livolant} deal with a finite
concentration of macroions that contribute in and of themselves to the
equation of state, i.e., to the osmotic pressure and osmotic
coefficient, the simple condenstaion picture appears to be too drastic
an idealization.  

%Quantitative discrepancies between the results
%derived within this limit and the consequences of the PB cell model
%(involving finite macroion concentration) as well as experimental data
%corroborate this view completely.

\vspace{1cm}
\noindent
% Extended condensation theory
%The alternative solution of attempting to formulate'extended' 
%condensation theories which have better control over effects such as 
%finite-radius effects, has been  pursued by quite a number of 
%researchers \cite{LeBret, Stigter, Belloni, Holm}. 
%To  a large extent these approaches can be described as attempts
%to obatin from an analysis of the nonlinear double layer equations or 
%from numerical simulation involving explicit ions, 
%a more general definition/identification of a 
%condensed layer of counterions near the polyectrolyte than is available
%in the 'simple' theory. These
%approaches involve a distinction between unbound and bound fractions
%of counterions which is perhaps less clear than  the distinction made 
%in 'simple' 
%theory, and the concept of condensation is largely operational, yet at the same time 
%more flexible than in 'simple' condensation theory:    

There have been many elegant attempts to extend condensation theory
\cite{LeBret, Stigter, Belloni, Holm}.  In essence these are all
versions of the Poisson Boltzmann equation solved under various
conditions including finite radius, explicit ion simulation, condensed
layers, etc.  These models create a distinction between bound and
unbound ions or speak of condensation shells within the context of the
continuous distributions predicted.

{\it (i)} Many have been tempted to identify, via the cell model
analysis, the fraction $f=1-1/Q_{o}$ of counterion residing in the
shell $a<r<R_{M}$ with the condensed fraction, and so implicitly to
accept that the condensed layer is a spatially extended object with a
peculiar sensitivity to changes in, e.g., concentration of macroions. 
Generalizations of this approach have been considered where one
identifies a length scale similar to $R_{M}$ and the accompanying
'condensation shell' from an analysis of the counterion distribution
function versus linear \cite{Manning98} or logaritmic \cite{Belloni,
Holm} radial distance.  They provide operational definitions of a
'condensed layer' but these definitions are almost a matter of
terminology.

%These approaches may be said to work in that they provide an 
%operational definition of a 'condensed layer', yet 
%in our opinion
%they seem to come very close to making the issue of condensation mostly 
%a terminological one, since these definitions are mostly made in terms of
%quantities that appear in the traditional double layer analysis or could
%easily have been identified via double layer analysis: 

Does such 'condensation' add anything essentially new either to P-B
theory or to simple line-charge condensation?

%The condensation picture adds nothing essentially new to the analysis not
%even the simplicity and elegance of the 'simple' picture. 

None of these approaches makes it completely clear that there is a
'Debye-H\'uckel' cloud of counterions in the outer shell
$R_{M}<r<R_{0}$, although at some distance from the polyelectrolyte
potentials must weaken to a degree that a D-H description becomes
accurate .  

{\it (ii)} It has been argued recently \cite{Holm}, that the fact that
in the infinite-dilution limit the contact density $n(a)$ reaches a
constant limiting value $(1/e b\pi a^{2})(Q-1)^{2}/(2Q)$ leads one to
expect the 'existence of a close layer that cannot be diluted away',
in other words a condensed layer.  This would point to the conclusion
that counterion condensation has a reality going beyond that of being
a mere heuristic tool.  This is perhaps true but does not really prove
the point, in the sense that one can not even in this case distinguish
'condensation' from mere double layer properties.

Comparison of the measured osmotic pressures for DNA at various ionic
conditions suggests that at low salt concentrations the 'counterion
condensation' picture does not capture the main features of the data. 
The discrepancy experimental and theoretical values of the osmotic
coefficient, amounting to a factor of two can not be ignored.  The
cell-model in which the countercharge distribution is governed by the
Poisson - Boltzmann equation and where the macroion is modelled as a
cylinder of a finite radius produces a much better fit to data.  This
almost quantitative correspondence between the data and the
calculations brings back into focus the counterion atmosphere
without any need to invoke a 'condensation' behavior.

\vspace{1cm}
\noindent
\section{Acknowledgement:}
We would like to thank Joel Cohen, Sergey Leikin, Eric Raspaud and 
Don Rau for stimulating discussions and Eric Raspaud for generously 
providing us with data prior to publication.

%                          *******************************

\newpage

\vspace{5mm} 
\noindent 
{\bf Figure 1:} Osmotic coefficient, $\phi=n(R_{0})/n_{PO4}$ vs. 
mean counterion concentration $n_{PO4}$ ($c_{DNA}$ in molar units) and
the geometry of the cell model.  $n({R_{0}})$ is the counterion
density at the cell wall.  The charge density on the cylinder
corresponds to one elementary charge per $1.7 \AA$ contour length (or
Q=4.353).  Osmotic coefficients have been calculated for $a=10 \AA$. 
We have considered the consequences of allowing the counterions to
enter  the interior of the cylinder (dashed line), or of preventing
the counterions from entering the cylinder at all (solid line). 
The  osmotic pressure in the limit of vanishing $a$ (Manning condensation 
limit) is shown by the  dashed line.
The calculated osmotic coefficients increase as $c_{DNA}$ increases. 
The experimental results for DNA \cite{Livolant} ($\blacksquare$) 
in the salt independent regime of DNA concentrations show the same trend,
but the variation is slower than for any of the calculated data.  The
osmotic coefficient is somewhat sensitive to whether the cylinder is
hollow (dotted lines) or solid (solid lines).  Expectably when the
cylinder is hollow, the osmotic coefficient is smaller.  The
hollow/solid difference increases when $R_{0} \longrightarrow a$.  In
the dilute, $a/R_{0} \longrightarrow 0$, limit the computed osmotic
coefficient reaches the constant limiting value of $1/(2Q) \simeq .11$
of counterion condensation theory.
 
\vspace{5mm} \noindent {\bf Figure 2:} Osmotic pressure ${\pi}_{osm}$
versus the molar concentration of DN,A $c_{DNA}$ (upper graph) or the
radius of the cell $R_0$ (lower graph).  The charge density on the
cylinder with radius $a=10 \AA$ corresponds to one elementary charge
per $1.7 \AA$ contour length (or $Q=4.353$).  Hollow cylinder, dotted
line; solid cylinder, solid line.  The predicted osmotic pressure in the
limit of vanishing $a$ (counterion condensation limit) is shown with
a dashed line.  Experimental data are represented with symbols:
$\blacktriangledown$ osmotic pressure data at 2 mM salt
\cite{Livolant}, $\blacksquare$ osmotic pressure data at DNA
concentrations where salt does not matter, $\large \bullet$ osmotic
pressure at very large DNA concentrations where again there are no
salt concentration effects \cite{Helmut}.  The experimentally determined pressures
for very small values of $c_{DNA}$ can not be explained by the
electrostatic cell model studied here.  For $c_{DNA}$ in the range
$0.1-0.5 M$, the calculated pressures determined from the
electrostatic cell model seems to capture the trend in the
experimental data but the calculated pressures are slightly larger
than the experimentally determined pressures, irrespective of
solid-cylinder (solid line) or hollow-cylinder (dashed line)
assumptions.  When $c_{DNA}$ approaches $1$ M , we find that the
calculated pressures and the experimentally determined pressures are
not very different in magnitude, but the calculated variation of
pressure with concentration does not fit the trend in the experimental
data.
 
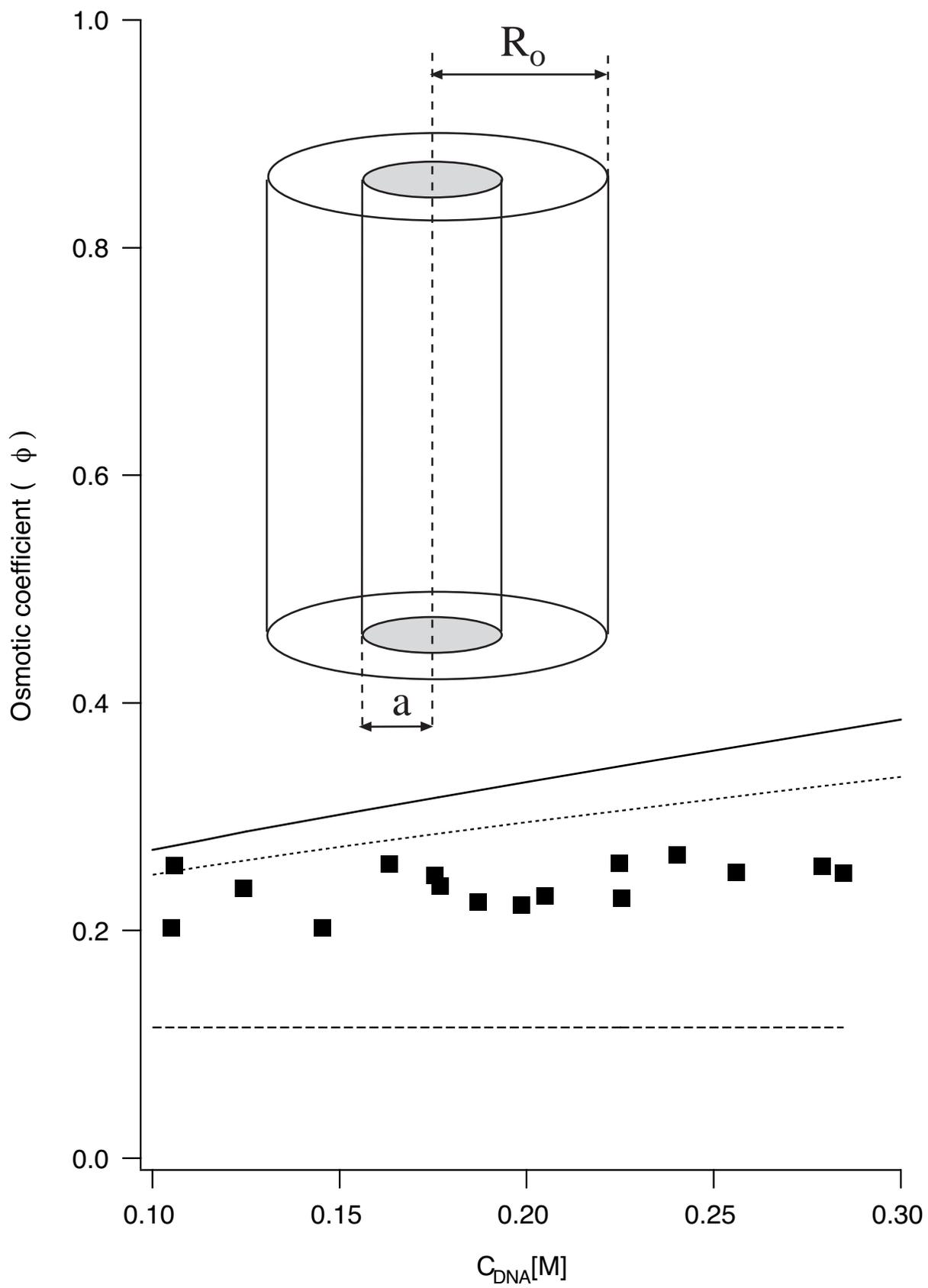
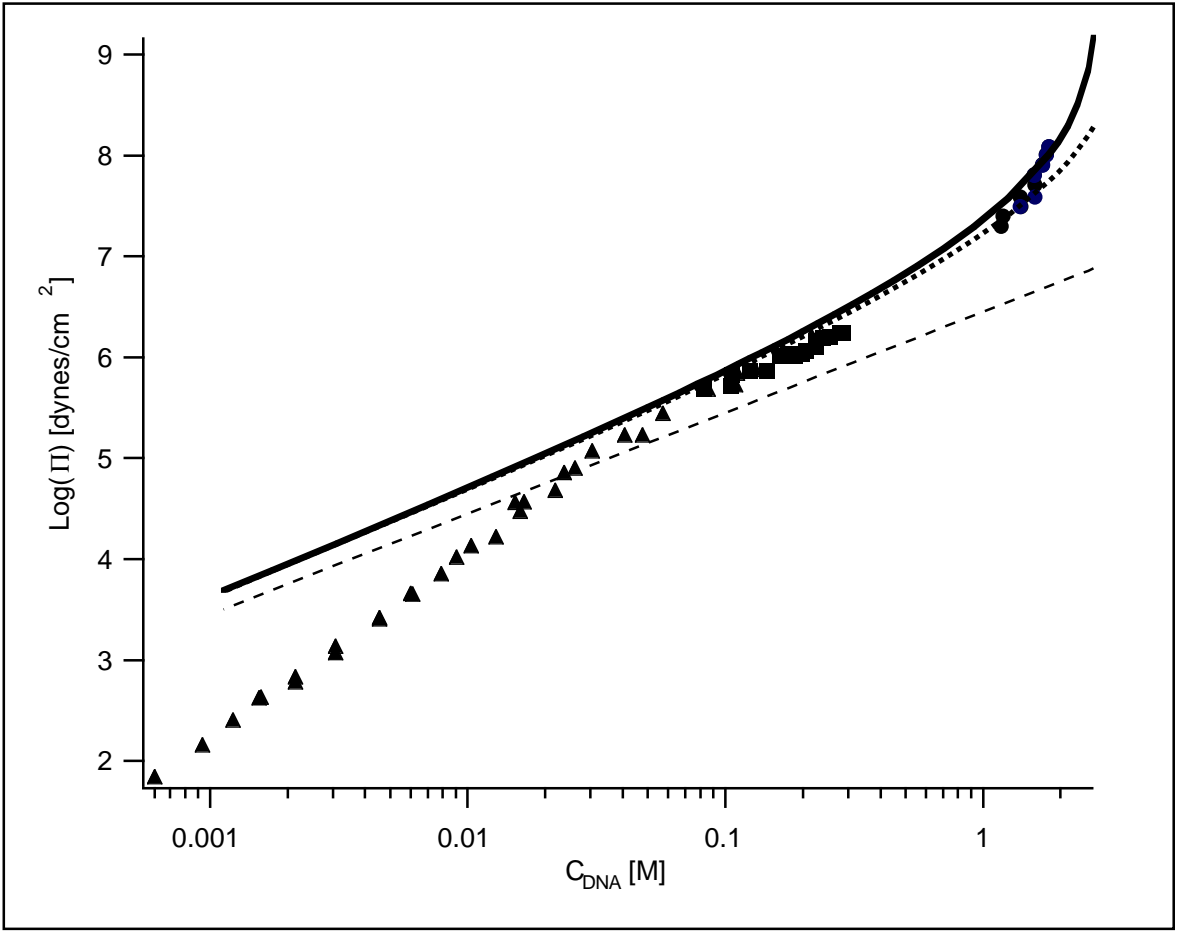
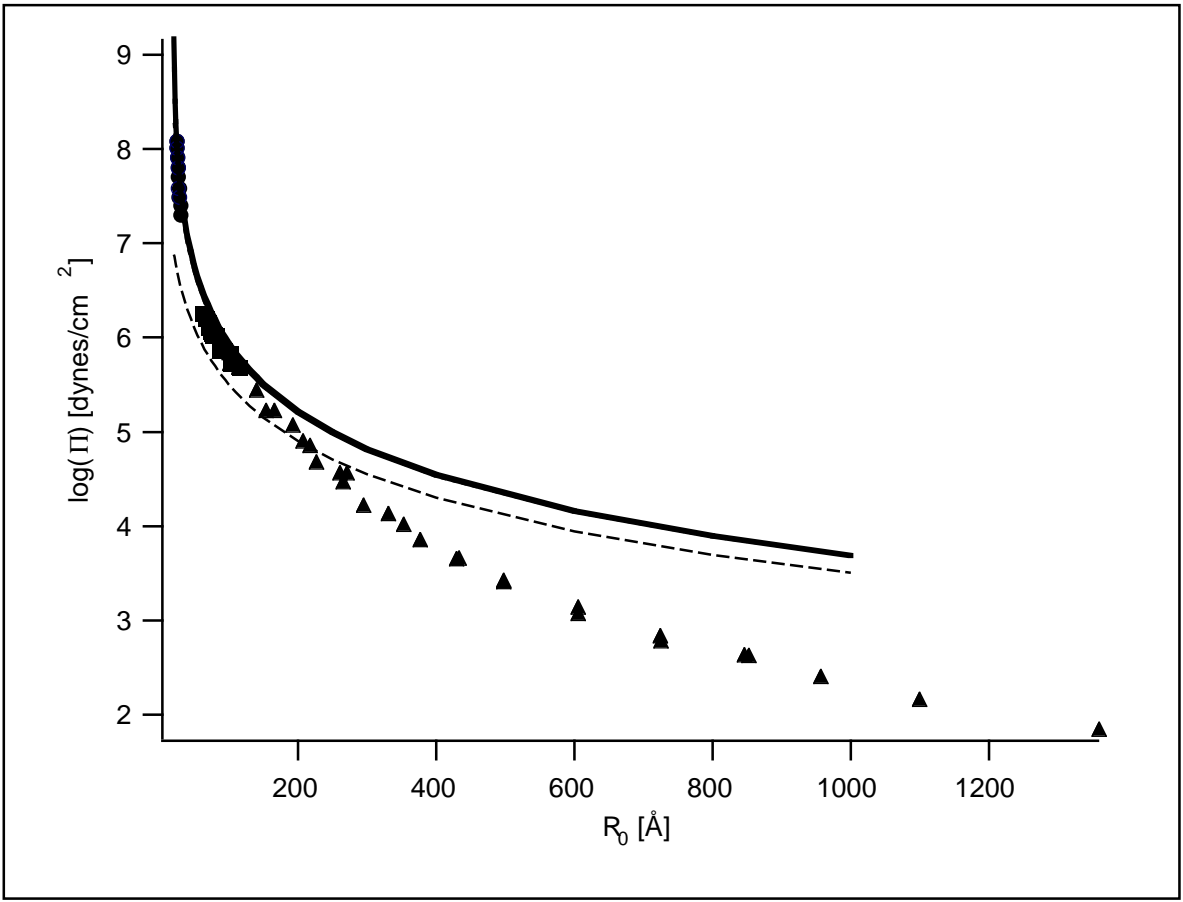
\end{document}